\documentclass[twocolumn,showpacs,amsmath,amssymb,prb,%
               superscriptaddress
              ]{revtex4}

\usepackage[T1]{fontenc}
\usepackage{times}
\usepackage{graphicx}
\usepackage{wasysym}
\usepackage{bm}

\begin{document}

\title{Strong diamagnetic response of metamaterials}

\author{E.~N.~Economou}
\affiliation{Dept.~of Physics,
             University of Crete, Heraklion, Crete, Greece}
\affiliation{Institute of Electronic Structure and Laser (IESL) --
             Foundation for Research and Technology-Hellas (FORTH),
             Heraklion, Crete, Greece}

\author{Th.~Koschny}
\affiliation{Ames Laboratory and Dept.~of Physics and Astronomy,
             Iowa State University, Ames, Iowa 50011, U.S.A.}
\affiliation{Institute of Electronic Structure and Laser (IESL) --
             Foundation for Research and Technology-Hellas (FORTH),
             Heraklion, Crete, Greece}

\author{C.~M.~Soukoulis}
\affiliation{Ames Laboratory and Dept.~of Physics and Astronomy,
             Iowa State University, Ames, Iowa 50011, U.S.A.}
\affiliation{Institute of Electronic Structure and Laser (IESL) --
             Foundation for Research and Technology-Hellas (FORTH),
             Heraklion, Crete, Greece}
\affiliation{Dept.~of Materials Science and Technology, 
             University of Crete, Heraklion, Crete, Greece}

\date{\today}

\begin{abstract}
We demonstrate that there is a strong diamagnetic response of metamaterials,
consisting of open or closed split ring resonators (SRRs). Detailed numerical
work shows that for densely packed SRRs the magnetic permeability,
$\mu(\omega)$, does not approach unity, as expected for frequencies lower and
higher than the resonance frequency, $\omega_0$.  Below $\omega_0$,
$\mu(\omega)$ gives values ranging from $0.9$ to $0.6$ depending of the width of
the metallic ring, while above $\omega_0$, $\mu(\omega)$ is close to $0.5$.
Closed rings have $\mu\approx 0.5$ over a wide frequency range independently of
the width of the ring. A simple model that uses the inner and outer current
loop of the SRRs can easily explain theoretically this strong diamagnetic
response, which can be used in magnetic levitation.
\end{abstract}



\maketitle

Several types of regular materials exhibit diamagnetic behavior characterized
by an induced magnetic moment opposite to the external magnetic field so that
their magnetic susceptibility $\chi$ is negative. 
Most of the natural diamagnetic materials have very low values of $\chi$, 
of the order of $10^{-6}$. The largest known diamagnetic value of a natural
material is that of pyrolytic graphite which is equal to 
$-4.5\times 10^{-4}$. 
A strong diamagnetic response is very important since it can be used in 
magnetic levitation.
There are two cases of strong diamagnetic response: 
One is that of the superconducting state which shows $\chi=-1$ (or weaker in 
the mixed state) practically over all frequencies. 
The other is that of magnetic metamaterials to be examined in this work.

Over the last seven years artificial materials have been designed and
fabricated consisting of an assembly of units (sometimes called magnetic
"atoms") which exhibit a resonant response to electromagnetic field of
appropriate polarization driving thus the magnetic permeability $\mu=1+\chi$ to 
negative values for a narrow frequency range just above the resonance frequency 
\cite{Smith,Soukoulis-1,Soukoulis-2}.
Given the fact that the magnetic response is in general a weak $v^2/c^2$ effect
\cite{Landau}, it is not unreasonable to expect that $\mu$ would approach unity
away from the resonance. 
However, as it can be seen from Fig.~1, this is not the case. 
Unusually large values of $|\chi|$ appear leading to values of $\mu$ as low 
as $0.5$.
More specifically, there is a frequency regime below the resonance (Fig.~1a)
extending over at least two orders of magnitude where $\mu$ stays clearly lower
than one; the larger the width $w$ of the ring (see caption of Fig.~1), the
smaller the value of $\mu$ is and the wider this frequency regime is. 
On the other side of the resonance, there is again a plateau where $\mu$ has 
a value of about $0.5$ for the chosen values of the parameters. 
This value is independent of the width $w$, and it is practically equal to the 
value of $\mu$ appearing if the ring is closed (See Fig.~1b showing a constant 
$\mu$ clearly less than one over a very broad frequency range extending over 
$3\frac{1}{2}$ orders of magnitude). 
We point out that the diamagnetic strength shown in Fig.~1b or 1a is comparable 
to that of the superconducting state.

\begin{figure}
 \begin{center}
  \includegraphics[width=4.25cm]{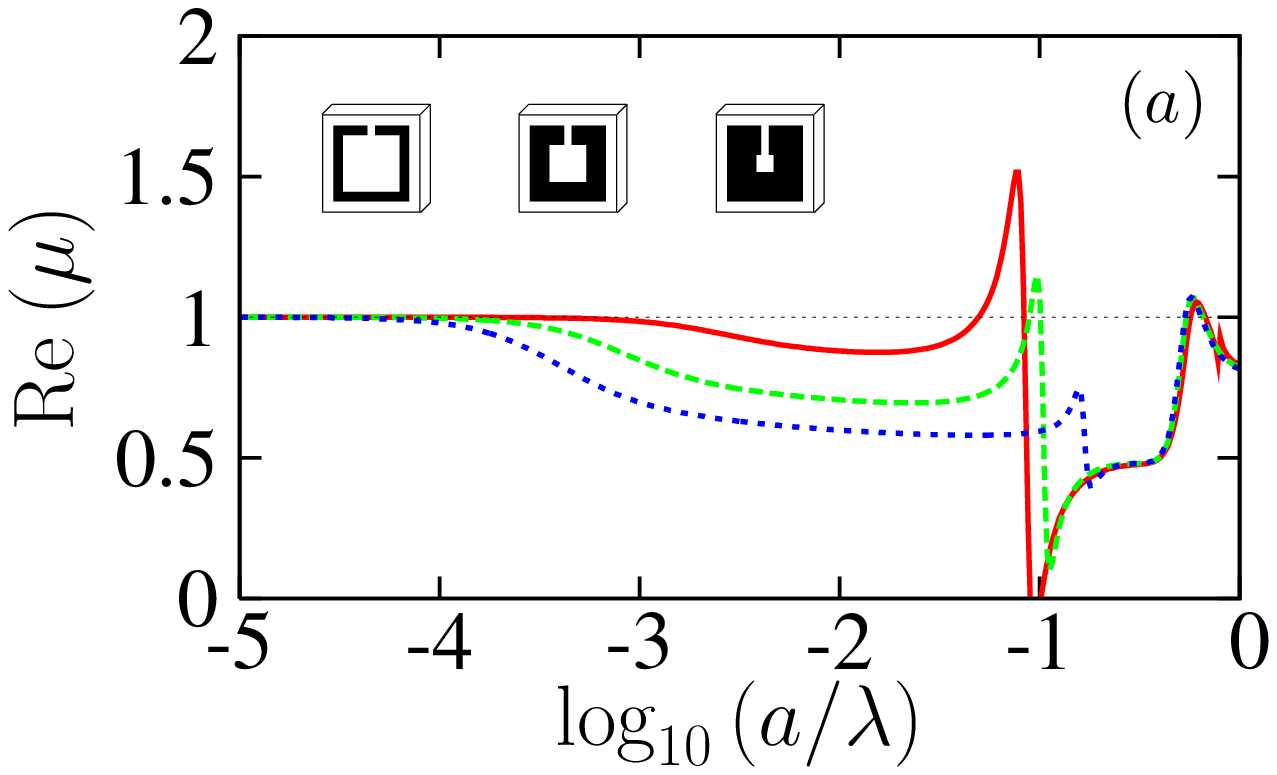}
  \includegraphics[width=4.25cm]{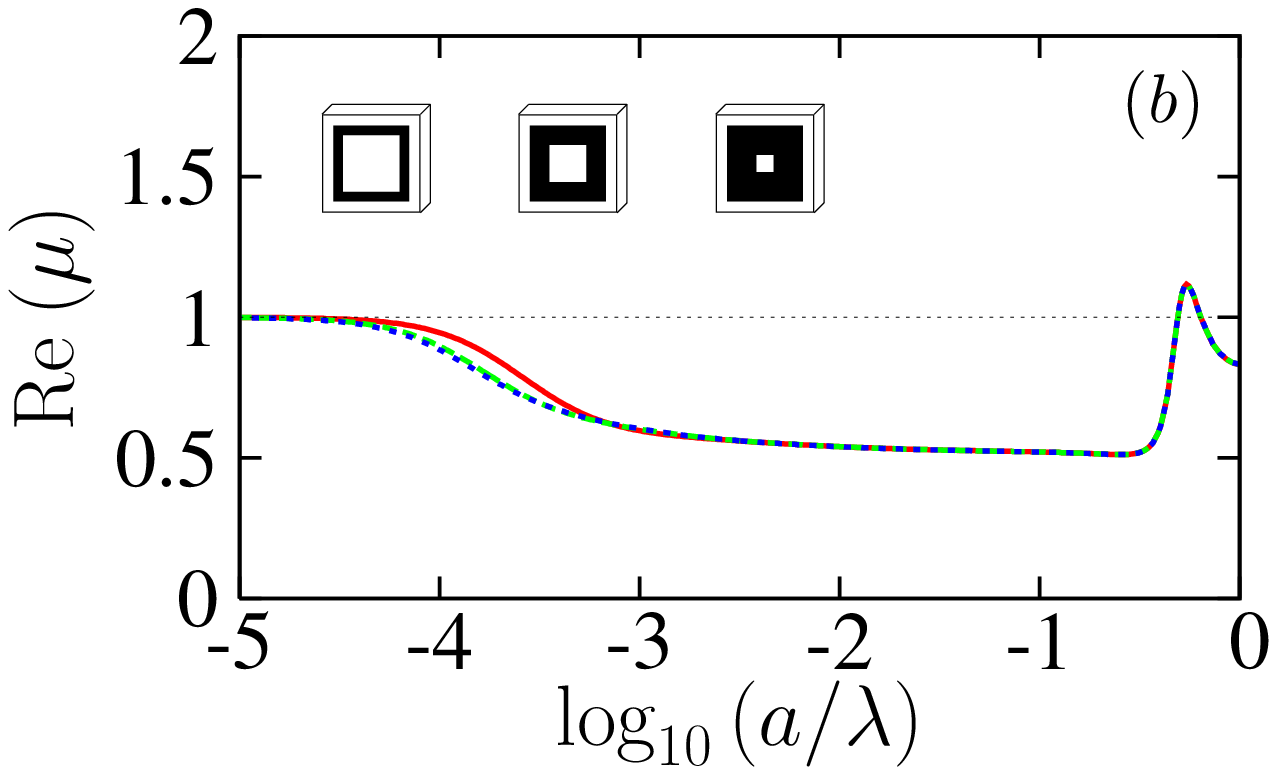} \\
  \includegraphics[width=4.25cm]{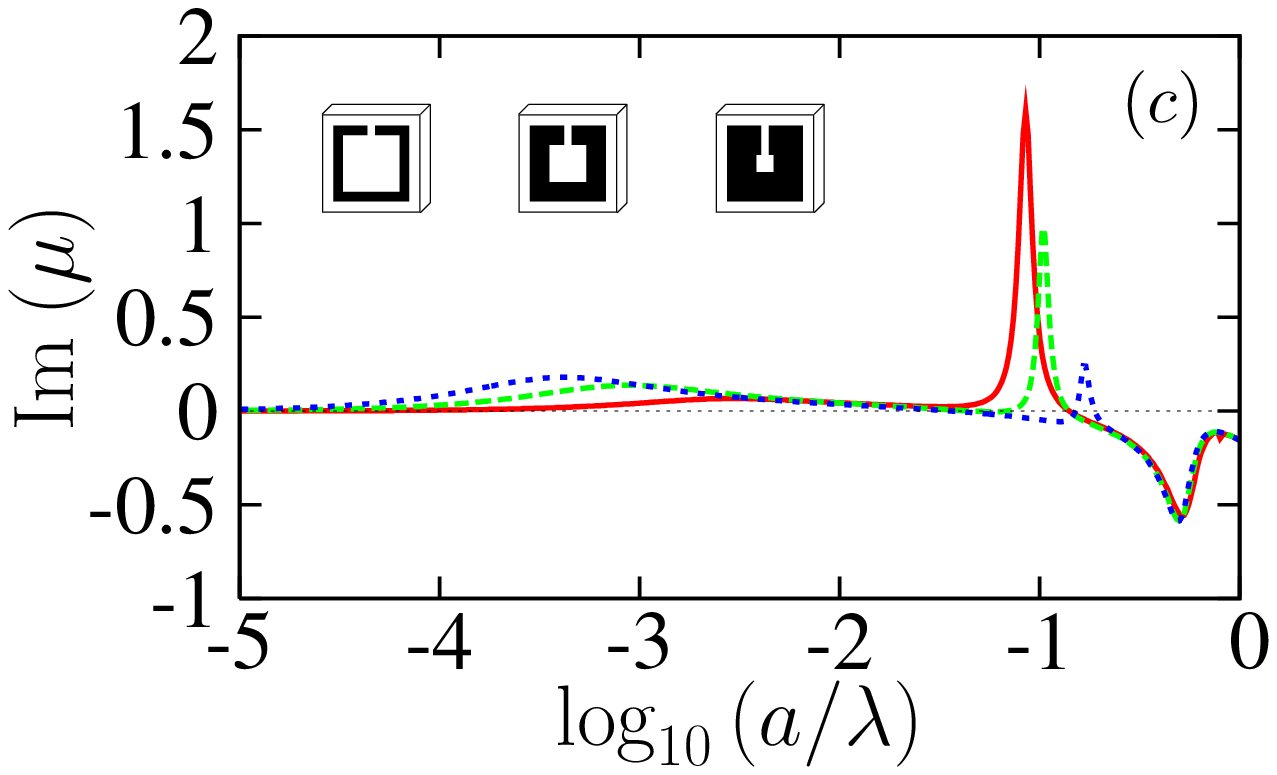}
  \includegraphics[width=4.25cm]{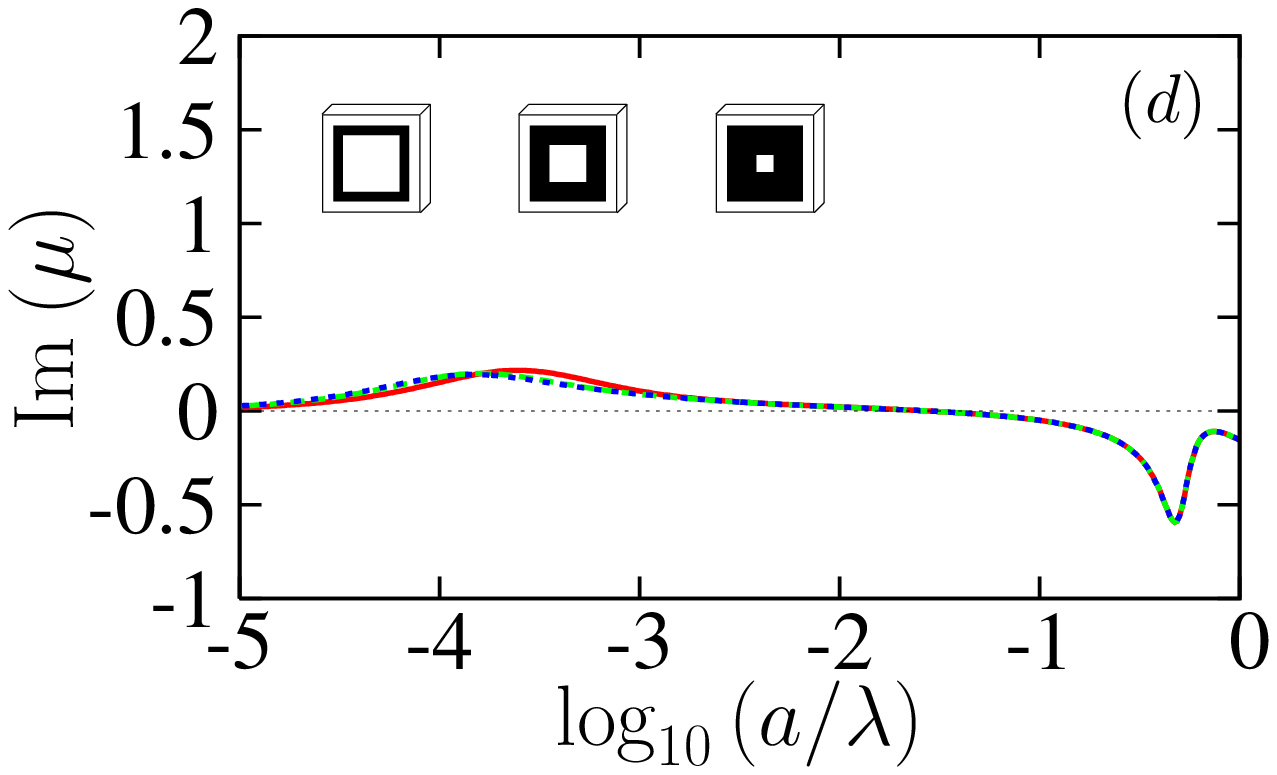}
 \end{center}
 \caption{%
 (Color online)
  $\mathrm{Re}\ \mu$ vs.~$\log_{10} (a/\lambda)$ obtained by simulations for a
  periodic assembly of open (a) or closed (b) square metallic rings.
  The corresponding $\mathrm{Im}\ \mu$ vs.~$\log_{10} (a/\lambda)$ are shown
  in panels (c) and (d), respectively.
  The simulations have been done using the commercial software packet Comsol
  MultiPhysics, which is essentially a finite-element method frequency domain
  solver.  The complex current distribution was obtained directly from the
  simulated field distributions and, averaged over the unit cell, gave the
  magnetization 
  (This averaging produces unphysical results when the wavelength $\lambda$
   becomes smaller then the unit cell size as evidenced by $\mathrm{Im}\ \mu < 0$.). 
  Using (4) together with the incident field $H_0$ we found the
  effective permeability shown.
  Unit cell size is $a\times a\times a_\bot = 10\times 10\times 2\ \mathrm{\mu m}^3$.
  Size of the ring is $\ell\times \ell = 8\times 8\ \mathrm{\mu m}^2$;
  its cross-section is $d\times w$, where the depth $d=400\mathrm{nm}$ and
  the width $w$ takes the values $1\mathrm{\mu m}$ (red, solid curve), 
  $2\mathrm{\mu m}$ (green, dashed curve), and $3\mathrm{\mu m}$ (blue, dotted curve).
  The metal is Drude-model gold with experimental values for the plasma frequency, 
  $\omega_p=2\pi\times 2184\,\mathrm{THz}$, 
  and collision frequency, $\omega_\tau=2\pi\times 6.5\,\mathrm{THz}$ \cite{Ordal}.
 }
 \label{fig:1}
\end{figure}

While the resonance region has been studied extensively, the other diamagnetic
regimes have received little attention\cite{Soukoulis-3,Hu,Wood}. 
In this letter we focus on these regimes and we explain their main unexpected 
features. 
We start with the approach of Gorkunov {\it et al.}\cite{Gorkunov}, according to which 
its inductance $L$, its capacitance $C$, and its resistance $R$ characterize 
each open ring. 
For closed rings there is no capacitance; the mutual inductances $L_{nm}$ among 
the rings (which are periodically placed to form a lattice) are also taken into 
account. 
The final result for $\mu$ and for open rings is 
\begin{equation}
 \mu = 1 - \frac{B \omega^2}{\omega^2 - \Omega_m^2 + i\omega \Gamma_m^{}}
\end{equation}
where $B = \mu_0 n A^2 Q^2 /L$, 
$n$ is the concentration of rings, $n = 1/(a^2 a_\bot)$, 
$A$ is the area of each ring, $A \approx (\ell-w)^2$,
$Q^2/L = (L_\mathrm{eff}+\frac{1}{3}\mu_0 n A^2)^{-1}$,  
$L_\mathrm{eff} = L+\sum_mL_{nm}$,
$\Omega_m^2 = Q^2\omega_0^2$, 
$\Gamma_m^{} = Q^2\gamma_0^{}$, 
$\omega_0^{} = 1/\sqrt{LC}$, 
and $\gamma_0^{} = R/L$.
For closed rings $\Omega_m^{} = 0$.
Notice that $\mu_0 n A^2 = (\mu_0 A/a_\bot) f_1^{}$ 
where $\mu_0 A/a_\bot$ is $L_\mathrm{eff}$ in the solenoid limit, 
and $f_1^{} = A/a^2$ is the two dimensional filling factor.
Thus in the limit of close-packed rings
\begin{equation}
 B \approx \frac{f_1^{}}{f_2^{}+f_1^{}/3}
\end{equation}
where $f_2^{}$ is smaller than but close to one.

\begin{figure}
 \begin{center}
  \includegraphics[width=4.25cm]{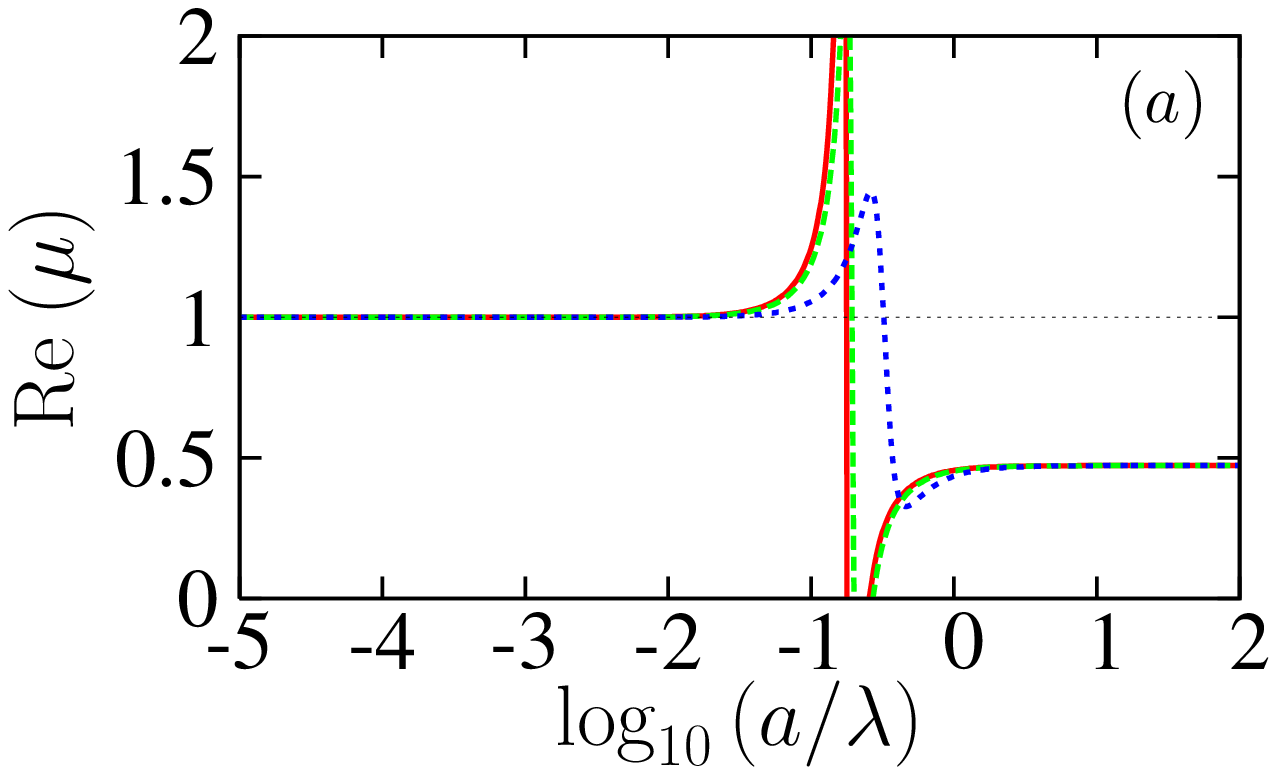}
  \includegraphics[width=4.25cm]{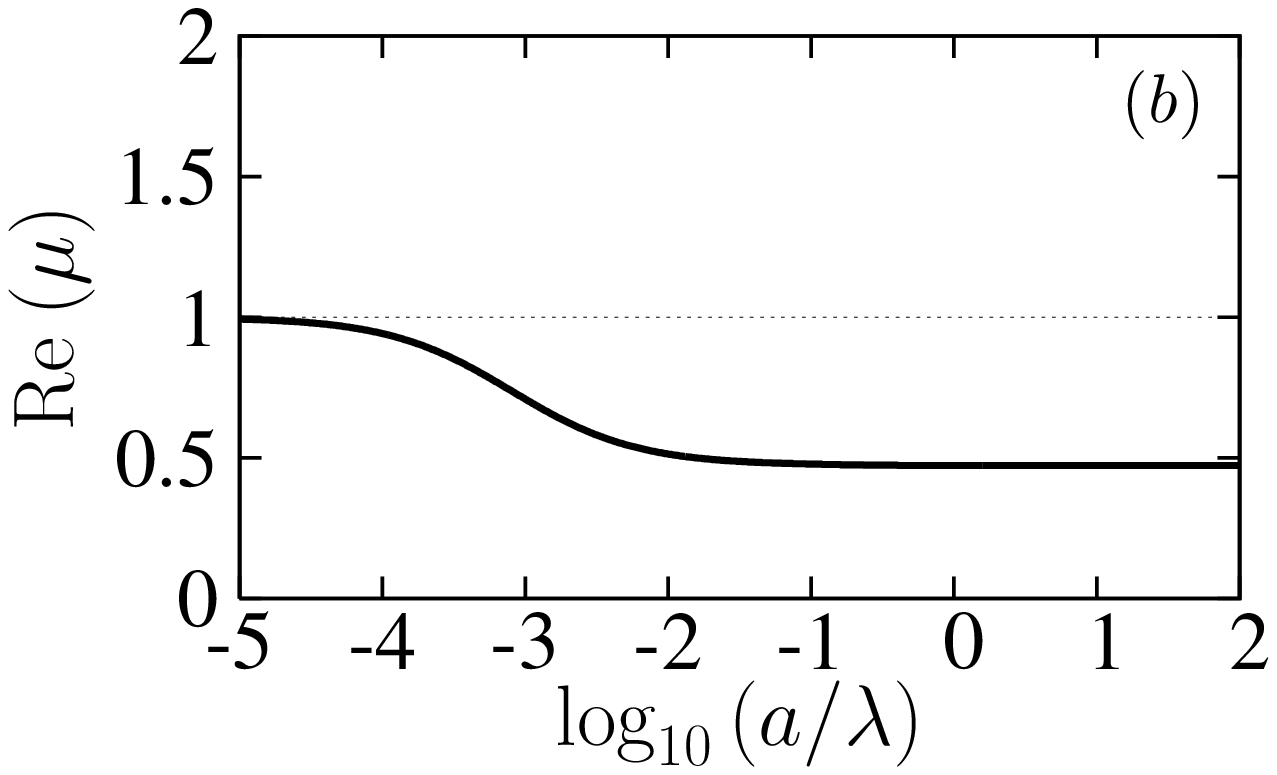} \\
  \includegraphics[width=4.25cm]{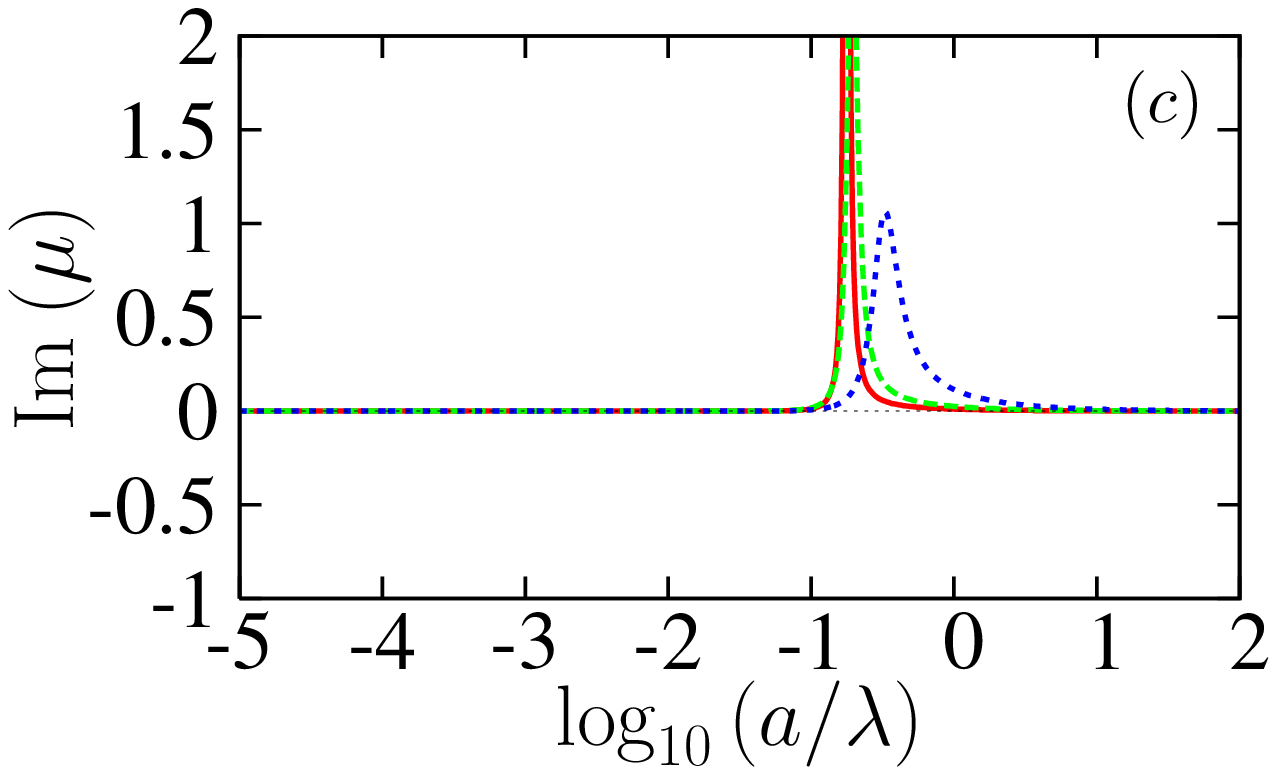}
  \includegraphics[width=4.25cm]{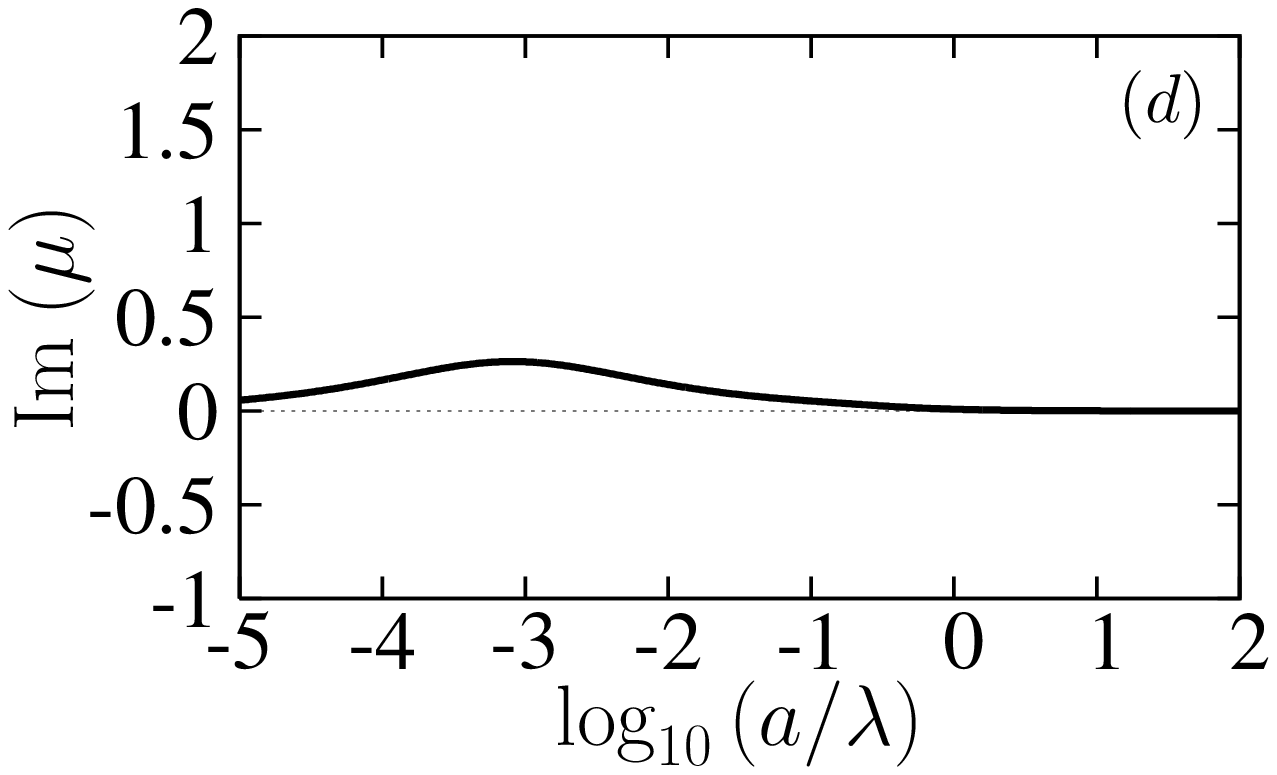}
 \end{center}
 \caption{%
 (Color online)
  $\mathrm{Re}\ \mu$ vs.~$\log_{10} (a/\lambda)$ according to Eq.~(1) for open
  rings (a) for the values of the width $w=1\mathrm{\mu m}$ (red, solid curve),
  $2\mathrm{\mu m}$ (green, dashed curve), and $3\mathrm{\mu m}$ (blue, dotted curve) 
  and for closed rings (b) of any width. 
  The corresponding $\mathrm{Im}\ \mu$ vs.~$\log_{10} (a/\lambda)$ is shown in (c) and (d),
  respectively.
  All parameters are as in Fig.~1.
 }
 \label{fig:2}
\end{figure}
In Fig.~2 we plot $\mu$ vs.~$\log_{10} (a/\lambda)$ according to Eq.~(1).
We have taken into account that $R$ and $L$ (to a lesser degree) depend 
on $\omega$ because of the skin effect\cite{Landau}. 
We see that the resonance region as well as the plateau above is 
reproduced fairly well, although the resonance is stronger and sharper 
than in Fig.~1a. 
Similarly, the closed ring result is in good agreement with the simulations 
data of Fig.~1b except of the final high frequency rise towards one, 
which we attribute to the fact that at this high frequency the wavelength $\lambda_m$
in the medium is comparable to twice the unit cell size; then the uniform field
assumption, on which the retrieval of $\mu$ from the simulation and Eq.~(1) is based, 
breaks down\cite{PEM}.

\begin{figure}[b]
 \centerline{\includegraphics[width=8.5cm]{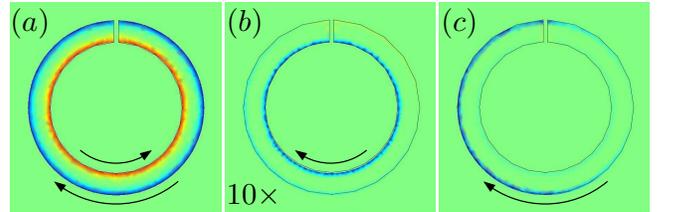}}
 \caption{%
 (Color online)
  Current distribution in the SRR ring 
  (the color/gray level is the azimuthal current density,
   the arrows indicate current direction at the edges)
  for the three regimes shown in Fig.~1 when the
  external magnetic field is normal to the plane of the rings and pointing out of it:
  (a) In the diamagnetic regime below resonance the current is confined near the
  outer and the inner edges flowing in opposite directions enclosing the metal filled area
  of the ring trace;
  (b) at resonance the current is mainly confined near the inner surface;
  (c) above resonance the current flows clockwise and essentially at 
  the outer edges of the metallic ring.
 }
 \label{fig:3}
\end{figure}
On the other hand, Eq.~(1) fails to reproduce the diamagnetic response below the resonance 
for the open ring case. 
To understand the reason behind this failure, we return to the simulations 
and we analyze the current distribution as shown in Fig.~3. 
In case (a), below the resonance, the opposite flowing currents at the two edges induce a 
magnetic field which cancels the external magnetic field inside the metallic wire. 
The wider the wire, the more extensive the diamagnetic volume is and the lower $\mu$ is. 
At resonance (case (b)), the strong clockwise current at the inner edge induces a strong field 
which in the inner area of the ring cancels or dominates over the external field producing thus 
the possibility of negative $\mu$. 
Finally, in the plateau above the resonance, the current flows clockwise near the outer edge 
inducing a field which cancels the external field over the area enclosed by the outer edge 
of the ring; this is the reason that the value of $\mu$ in this regime is independent of the 
metallic width $w$ (assuming constant $\ell$) and equal to the closed ring value, 
since in the latter case the current flow is as in case (c) over the 
entire diamagnetic regime.
The asymmetry of the circular current in case (c) is due to the superposition of growing 
linear currents caused by a electric cut-wires resonance at higher frequency.

\begin{figure}
 \centerline{\includegraphics[width=7cm]{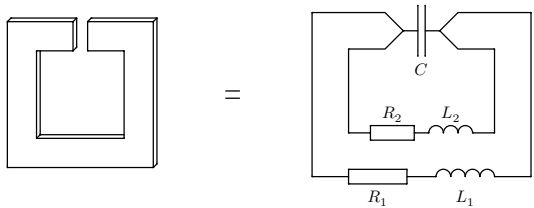}}
 \caption{%
  Equivalent circuit for an open ring.
  Loop 1 (2) represents the current path roughly one skin depth wide along 
  the outer (inner) edge of the ring.
  The two loops have a mutual inductance $L_{12}=L_{21}$. 
 }
 \label{fig:4}
\end{figure}
To model this more complicated behavior of the open ring, we introduce a two
loop representation of it sharing the single capacitance of the gap as in Fig.~4.
By defining the effective inductances 
$\tilde{L}_1 = L_1 + \sum_m L_{1,nm}$, 
$\tilde{L}_2 = L_2 + \sum_m L_{2,nm}$,
$\tilde{L}_{12} = \tilde{L}_{21} = L_{12} + \sum_m L_{12,nm}$,
we find using Kirchhoff's circuit equations the following matrix relation 
connecting the currents $I_1$ and $I_2$ in each loop to the external magnetic field $H_0$
\begin{equation}
\left(\begin{array}{cc} Z_{11} & Z_{12} \\ Z_{21} & Z_{22} \end{array}\right)
\left(\begin{array}{c} I_{1} \\ I_{2} \end{array}\right)
\ = \
i\omega \mu_0 H_0
\left(\begin{array}{c} A_{1} \\ A_{2} \end{array}\right)
\end{equation}
where $Z_{ii} = -i\omega\tilde{L}_{i} -(i\omega C)^{-1} +R_i$, $i=1,2$,
$Z_{12}=Z_{21} = -i\omega\tilde{L}_{12} -(i\omega C)^{-1}$, 
and $A_1$ $(A_2)$ is the area enclosed by the outer (inner) edge of the ring.
The permeability $\mu$ is given, following Ref.~6 by 
\begin{equation}
 \mu = \frac{\displaystyle 1+\frac{2}{3}\frac{M}{H_0}}
            {\displaystyle 1-\frac{1}{3}\frac{M}{H_0}}
\end{equation}
where the average magnetization $M$ is equal to $n(A_1 I_1 + A_2 I_2)$. 
In what follows we shall employ the solenoid approximation 
(which is good for close packed rings, i.e.~small $a_\bot$) 
in order to avoid the tedious lattice sums involved in the definition of 
$\tilde{L}_1$, $\tilde{L}_1$, and $\tilde{L}_{12}$,
which just renormalize parameters 
but do not qualitatively alter the physical behavior.
According to this approximation we have 
$\tilde{L}_1\approx \mu_0 A_1/a_\bot$,
$\tilde{L}_2\approx \mu_0 A_2/a_\bot$,
and $\tilde{L}_{12}=\tilde{L}_{21}$, $\tilde{L}_{12}\approx \mu_0 A_2/a_\bot$.

\begin{figure}
 \begin{center}
  \includegraphics[width=4.25cm]{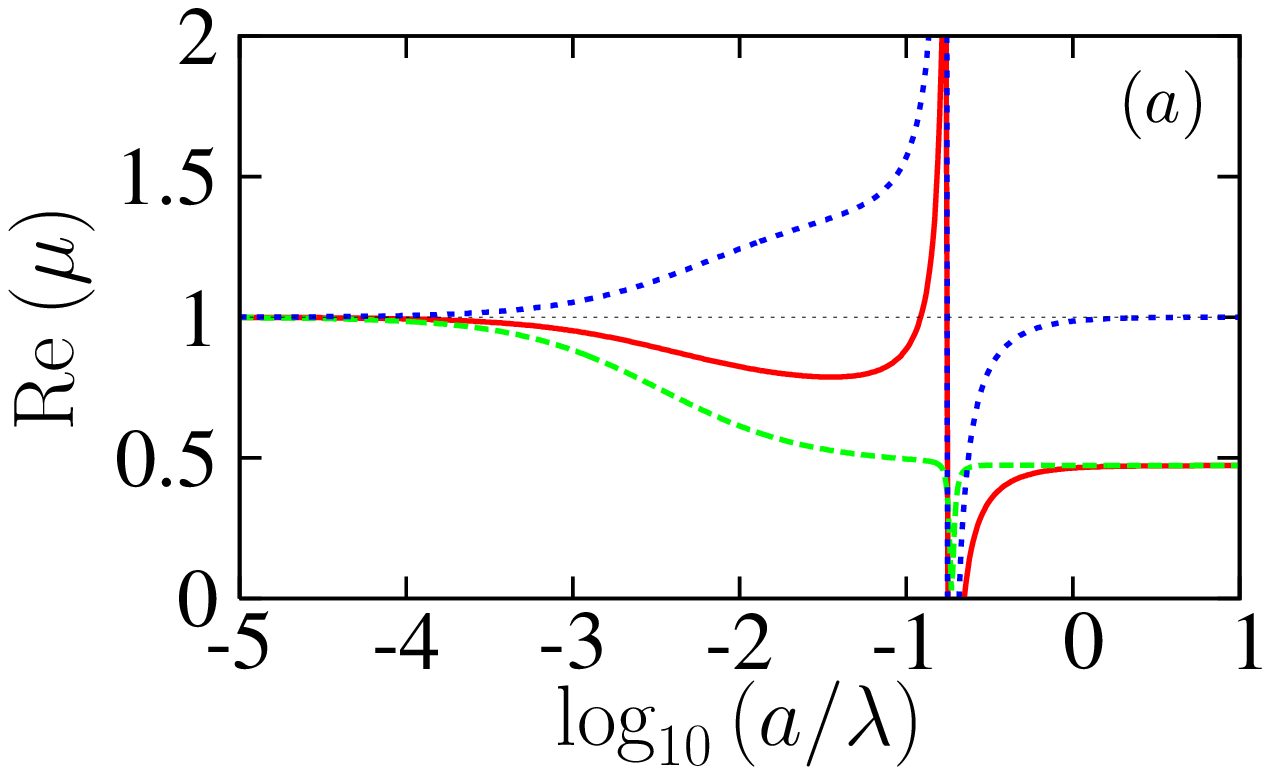}
  \includegraphics[width=4.25cm]{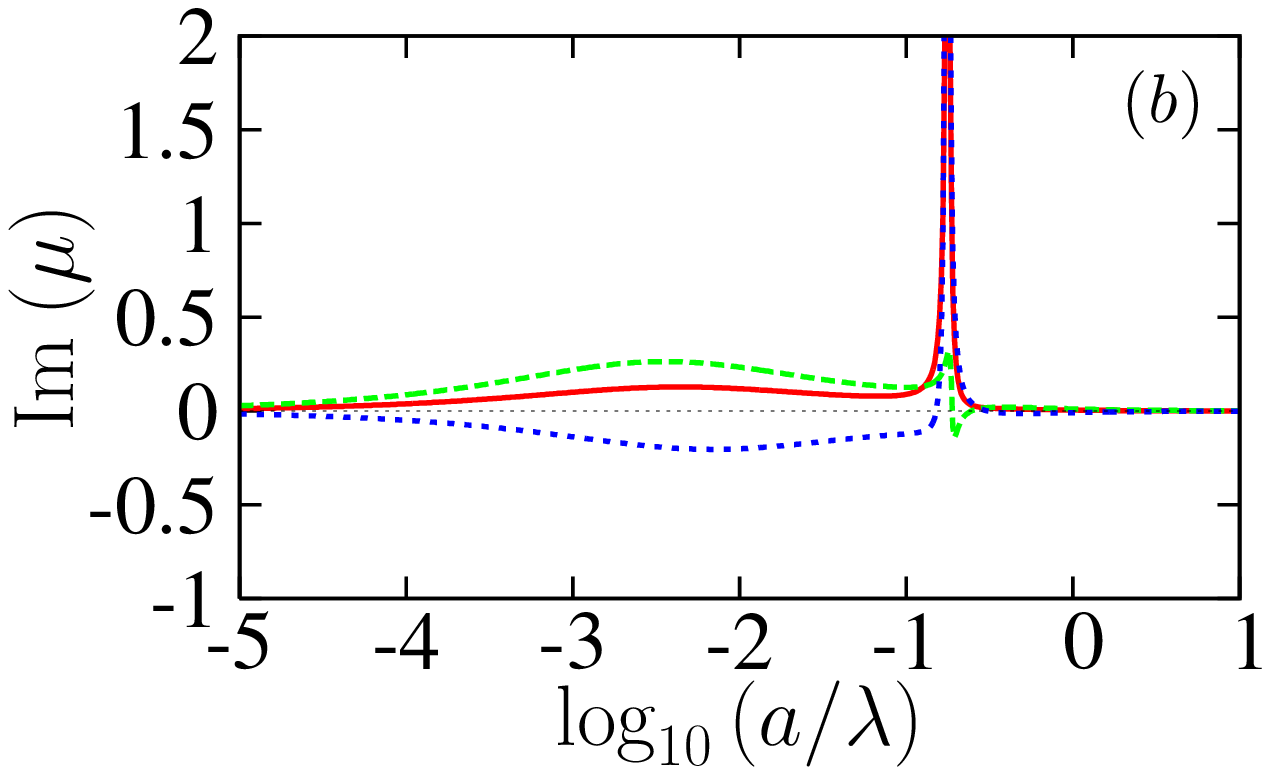} \\
  \includegraphics[width=4.25cm]{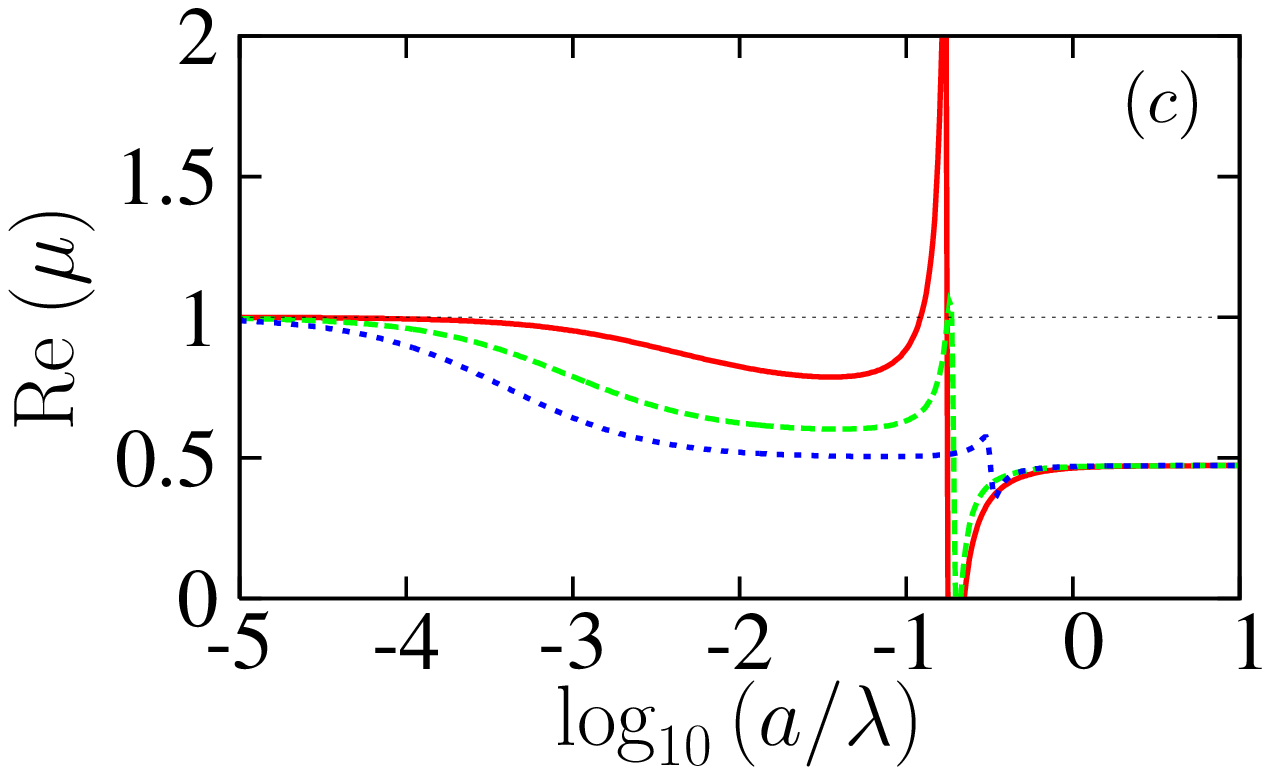}
  \includegraphics[width=4.25cm]{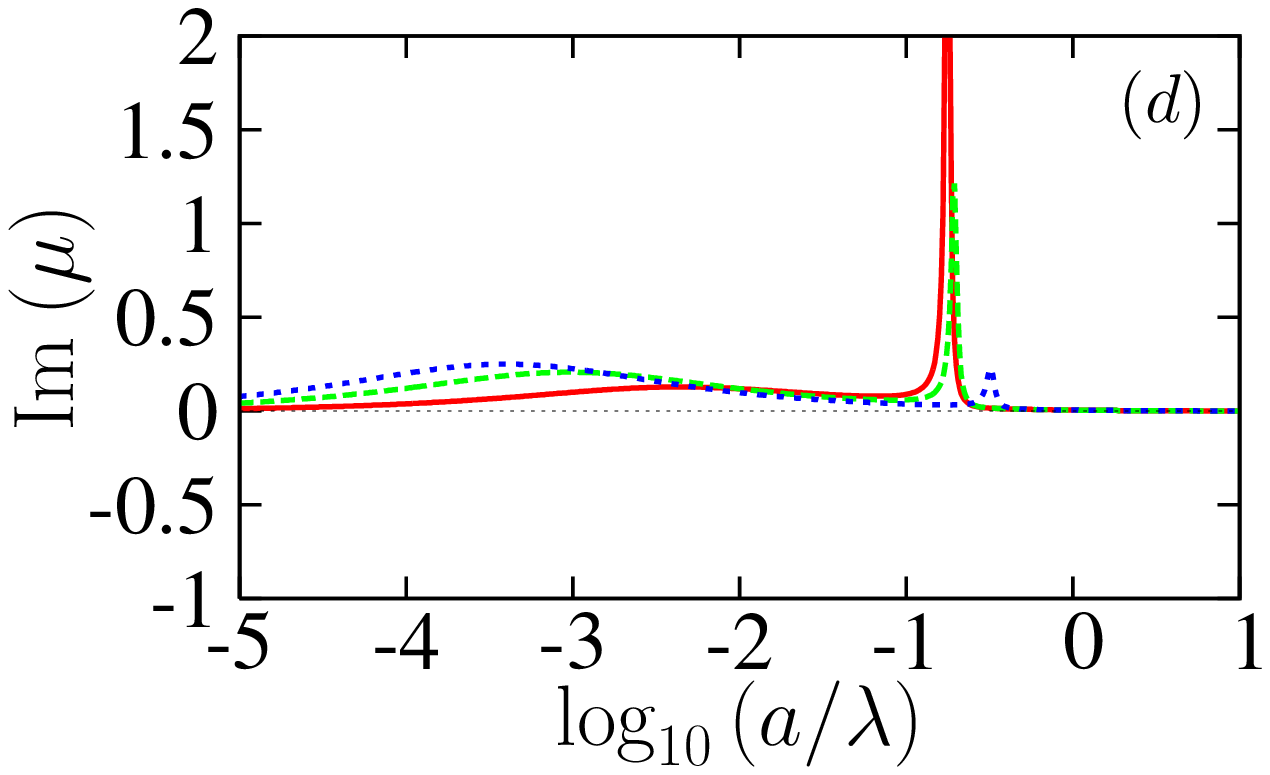}
 \end{center}
 \caption{%
 (Color online)
  (a) $\mathrm{Re}\ \mu$ vs.~$\log_{10} (a/\lambda)$ for the open SRR
  rings with $w=1\mathrm{\mu m}$ (red, solid curve) together with the contribution 
  of inner (blue, dotted curve) and the outer (green, dashed curve) edges,
  (b) shows the corresponding $\mathrm{Im}\ \mu$ vs.~$\log_{10} (a/\lambda)$ 
  (red, solid curve) together with the contribution
  of inner (blue, dotted curve) and the outer (green, dashed curve) edges.
  (c) and (d) show $\mathrm{Re}\ \mu$ vs.~$\log_{10} (a/\lambda)$ and 
  $\mathrm{Im}\ \mu$ vs.~$\log_{10} (a/\lambda)$, respectively, for the 
  three cases discussed in Fig. 1a with ring widths
  $1\mathrm{\mu m}$ (red, solid curve), $2\mathrm{\mu m}$ (green, dashed curve),
  and $3\mathrm{\mu m}$ (blue, dotted curve).
  These results were obtained according to Eqs (3) and (4).
 }
 \label{fig:5}
\end{figure}

In Figs.~5a,b we plot $\mu(\omega)$ together with the corresponding 
contributions from the outer and inner loops, 
while in Figs.~5c,d we plot $\mu(\omega)$ for the three cases shown 
in Fig.~1a according to Eqs.~(3) and (4).
We see from Fig.~5a that the diamagnetic contribution below resonance is coming
from the outer loop, which prevails over the contribution of the inner loop,
which is paramagnetic. 
At resonance the inner loop dominates, while the outer
loop give a negative peak there. 
Finally, above resonance the outer loop gives
a constant diamagnetic contribution, while the inner loop contribution
approaches smoothly the value $\mu=1$. 
The $\mathrm{Im}\ \mu(\omega)$ (Fig.~5b) confirms the predominance of
the inner loop at the resonance frequency. 
The contribution of the inner loop to $\mathrm{Im}\ \mu(\omega)$ in the 
diamagnetic region below resonance is negative, while the total 
$\mathrm{Im}\ \mu(\omega)$ is, of course, positive. 
This indicates that the outer loop feeds the inner loop with energy there. 
For comparison, we calculated the total $\mu(\omega)$ within the two-loop model
for open rings and for the three cases shown in Fig.~1a. 
The result is shown in Fig.~5c and it is in good agreement with the simulations 
presented in Fig.~1a.
We calculated also, within the two loop model, $\mu(\omega)$ for closed rings
and we found that it is constant and negative and equal to the high frequency
asymptotic value for open rings; this result is not plotted, since it
practically coincides with those in Figs.~1b and 2b.
Thus the two loop model reproduces the $\mu$ vs. $\omega$ dependence, as
determined by Femlab simulation, as well as all the features of the current
distributions in the various regimes, and provides a clear explanation for the
complicated diamagnetic behavior of the open rings.

In conclusion, the present work shows that there are five frequency regions, in
the magnetic response of open rings. 
This behavior can be understood in term of the outer and the inner current loop 
of each ring proposed here: 
The low frequency regime, $\omega < R/L$, exhibits practically no magnetic response,
$\mu=1$, because of the resistive damping as expected for most materials. 
The second frequency regime, $R/L < \omega < 1/\sqrt{LC}$, exhibits an unusually strong
diamagnetic behavior (even for metal volume filling ratios as low as 10\%) which is due
to the outer current loop, while the inner one makes a paramagnetic
contribution (the two currents are connected along the gap);
the net result depends on the difference $A_1 - A_2$ of the areas 
enclosed by these two loops. 
The third regime, $\omega\approx 1/\sqrt{LC}$, is the resonance region, 
which is dominated by the inner loop; 
this presents a design challenge, since what creates a strong diamagnetic background 
below the resonance frequency, namely a small area inner loop, makes the resonance weak. 
The fourth regime, $1/\sqrt{LC} < \omega < \pi c/a$, shows again a
very strong diamagnetic response and depends only on the outer loop, 
being proportional to $A_1$.
Finally, in the regime $\pi c/a < \omega$, the assumption of the wavelength $\lambda_m$
in the medium being much larger than the lattice constant $a$ breaks down and
both the averaging procedure employed in the simulations and
the description in terms of a simple effective electric circuit fail.

Work at the Ames Laboratory was supported by the Department of Energy (Basic
Energy Sciences) under Contract No. DE-AC02-07CH11358. This work was partially
supported by the AFOSR under MURI grant (FA9550-06-1-0337), by DARPA (Contract
No. MDA-972-01-2-0016), by Department of the Navy, Office of Naval Research
(Award No. N00014-07-1-0359), EU projects: Molecular Imaging
(LSHG-CT-2003-503259), Metamorphose and PHOREMOST, and by Greek Ministry of
Education Pythagoras project.

\bibliographystyle{apsrev}

\end{document}